# Integrating VR tours in online language learning: A design-based research study

**Roberto B. Figueroa Jr.**
University of the Philippines

**Insung Jung**
Education Research Institute, Seoul National University, Republic of Korea

This paper presents an investigation into the integration of virtual reality (VR) tours in online English lessons tailored for adult learners. The study utilised a design-based research approach to evaluate the effectiveness of VR tours in this context. It specifically examined the responses of adult learners to this instructional strategy by collecting data through surveys, observation notes and interviews with four learners in Japan and five learners in France, most of whom completed 10 lessons over 4 months. The research findings highlight the effectiveness of VR tours in enhancing learner motivation. Additionally, they demonstrate that perceived learning outcomes are influenced not only by the immersive experience of spatial presence but also by the novelty of technological and scenery-related aspects within the VR environment, as well as factors related to lesson design and individual learner characteristics.

*Implications for practice or policy:*
- VR photo tours can serve as a powerful tool for language educators to increase the learning motivation of remote learners.
- The physical location of remote learners and the ergonomics of their devices plays a crucial role in their engagement levels with VR-based online learning activities.
- Aside from the technological novelty of VR, the inclusion of diverse and visually appealing scenery can further enhance motivation and interest among online learners.

*Keywords:* virtual reality, online learning, educational technology, design-based research, virtual tours, motivation

## Introduction

Motivation is widely acknowledged as a key factor that influences learner engagement and outcomes in online learning environments, where students are physically distant from their instructors and educational institutions (e.g., Barak et al., 2016; Laato et al., 2019; Zheng et al., 2015). A variety of studies have explored numerous factors that impact motivation, drawing insights from various theories of motivation, including goal theory, future self-guides, self-determination theory and interest theory. These influencing factors include enjoyment (Okada & Sheehy, 2020), internal locus of control, satisfaction, flow (Y. Lee & Choi, 2013), self-regulation (Reparaz et al., 2020; Shea & Bidjerano, 2010), situational interest (J. C. Y. Sun & Rueda, 2012) and the ideal future-self concept (Dornyei, 2019).

In the realm of virtual reality (VR)-based learning interventions, situational interest – which is generated by specific aspects of an event or object (Hidi & Renninger, 2006) – and the ideal future-self – representing an individual's desired possible self, encompassing their dreams and goals (Dornyei, 2019) – have been identified as particularly relevant. This significance arises from VR's ability to generate situational interest and facilitate the envisioning of the ideal future-self through its novelty and interactive experiences. Alongside these variables, immersive capability, which refers to the objective level of sensory realism a VR system delivers, and spatial presence, the feeling of being physically present in a virtual environment, have also been recognised as key factors influencing learner engagement and outcomes in VR-based learning (e.g., Blomstervik & Olsen, 2022; Figueroa, 2023).

Overall, VR-based interventions have the potential to improve learner experiences and outcomes by fostering motivation through enhanced situational interest, the development of one's ideal future self-image, immersive capability and spatial presence. This perspective is supported by extensive research in traditional, in-person classroom settings (Dhimolea et al., 2022; Frazier et al., 2021; Jin, 2021). However, empirical research in online and non-formal language learning settings, where sustained motivation is crucial, remains limited.

To address this research gap, the study explored the interplay of VR-related variables, including situational interest, ideal future-self and spatial presence, and their impact on language learning outcomes, without looking into immersive capability due to the use of low-cost VR spectacles as an only option. It analysed nine adult learners from Japan and France in 10 VR-based English lessons over 4 months, using a design-based research (DBR) approach to identify effective features of VR English lessons through continuous refinement, focusing on three key research questions (RQs):
- RQ1: How did adult online learners experience VR tour-based English lessons, and in what ways did these experiences reflect opportunities for enhancing learner engagement and lesson improvement?
- RQ2: What patterns of spatial presence, situational interest, ideal future-self, and learning outcomes emerged across multiple VR tour-based lessons, particularly in sustaining motivation?
- RQ3: How did individual learners differ in their experiences and motivation dynamics (spatial presence, situational interest, ideal future-self, and learning outcomes) within these VR-based English lessons?

**VR and motivation**

The capabilities of VR applications in teaching and learning have captured the interest of scholars from a wide range of fields, from the arts to the physical sciences. Deng et al. (2021) investigated the use of VR in animation art education, arguing that VR can significantly enhance this field. Their study found that a remarkable 96.25% of students preferred the VR experimental teaching system for animation art, highlighting its educational potential.

Another study, by Du et al. (2022), introduced a comprehensive model incorporating task-technology fit and usage satisfaction to analyse factors influencing teachers' intentions to continue using VR technology. Surveying 291 teachers from elementary and secondary schools in Jiangsu Province, China, the study revealed that teachers' perceived usefulness and ease of use of VR technology significantly impacted their satisfaction with its usage.

In the context of sports dance education, the effectiveness and acceptance of VR technology were also studied, revealing a high acceptance rate (88.6%). This underscores the growing interest in and acceptance of innovative teaching methods in physical education. In language education, particularly English instruction, VR technology is gaining popularity. Chen (2022) explored how VR could aid English learners in speech communication, finding that public speaking anxiety levels among students who used VR were reduced significantly.

Although varied outcomes have been reported, two key variables impacting motivation in VR-based education are highlighted: *immersive capability* – VR's power to offer three-dimensional, computer-generated experiences (Asakawa et al., 2020; Hoffman et al., 2023) and *spatial presence* – the feeling of being in a virtual environment (Figueroa, 2023; Lessiter et al., 2001). A study among Japanese students found a positive correlation between spatial presence, motivation and perceived learning (Figueroa et al., 2022).

Other studies have shown that spatial presence can enhance situational interest (Cheng, 2022; Figueroa, 2023; J. C. Y. Sun et al., 2023), a psychological state of engagement, and support the visualisation of one's ideal future-self, thereby fostering motivation. Situational interest combines affective reactions and cognitive functioning, facilitating attention and learning (Hidi & Renninger, 2006; Schiefele, 2009),and is linked to self-regulation, a key trait for successful online learners (Reparaz et al., 2020; Shea & Bidjerano, 2010).

Furthermore, VR enables individuals to visualise their ideal future selves, exploring various careers and lifestyles through simulations. This bridges the gap between current self-perception and future aspirations. The ideal future-self, essential for language learning motivation (Dörnyei, 2019), requires a vivid image to inspire motivated behaviour, as shown in several studies (J. S. Lee et al., 2021; Liu, 2021). VR's ability to simulate visually rich and realistic environments may facilitate the creation or reinforcement of the ideal future-self, leading to motivated behaviour.

**Research gap in VR studies**

The potential of VR-based interventions in enhancing motivation for language learning has been demonstrated (Alizadeh & Cowie, 2022). Studies have further indicated VR's advantage in providing situational context and eliciting emotional responses to facilitate complex learning (Calvert & Hume, 2022). VR provides an effective means for simulating and visualising abstract concepts that may be challenging or impossible to obtain in class (Fan et al., 2010). According to Cochrane and Farley (2017), immersive technologies like VR allow for new ways of interacting with real environments and transport users to virtual learning spaces beyond their physical location. This supports the idea that VR lets learners experience and interact with simulated environments as if they were physically present, without being limited by access or safety concerns (Hoffman et al., 2001).

However, there remains a significant gap in research on the long-term effects of these interventions, particularly in contrast to the numerous short-term, experimental studies that dominate the field. Although experimental studies provide valuable insights into immediate learning outcomes, they often lack the longitudinal perspective needed to assess sustained engagement and learning gains over time.

Many VR-based interventions in language learning are typically one-off studies or short-term implementations lasting a few weeks (Lin & Lan, 2015). These studies often focus on immediate improvements in motivation or immersion but fail to investigate how VR impacts learners' language acquisition over longer periods, such as an academic semester or year (Jensen & Konradsen, 2017). For instance, although VR has been shown to boost learners' situational interest and provide immersive experiences (Schott & Marshall, 2021), the sustainability of this motivation is less well understood. The limited duration of most studies, typically ranging from a few hours to a couple of weeks, restricts our understanding of how learners might navigate the learning curve, adapt to VR technologies and continue to develop language skills in extended VR-based educational environments.

To address this gap, future research needs to adopt longitudinal approaches, tracking learners' progress and engagement in VR-based language learning environments over several months or more. This would provide a more comprehensive understanding of how VR can contribute to sustained motivation and language acquisition, beyond the short-term novelty effect observed in many experimental studies.

Furthermore, integrating VR-based activities in one-on-one online sessions with adult learners is relatively new. Thus, DBR could provide a useful iterative and reflective framework for studying its effectiveness while investigating individual differences and challenges that learners encounter in each iteration. This approach allows for the creation of pedagogically sound, language-learning-specific VR-based lessons ensuring that interventions are both technologically innovative and pedagogically effective, thereby improving language learning outcomes.

**DBR and the presentation, practice and production method**

DBR focuses on iterative cycles of design, testing in real-world teaching and analysing outcomes to refine solutions while deriving theoretical and pedagogical insights (Collins et al., 2004). DBR is unique for its long-term, in-context application, emphasising collaboration between researchers and practitioners to link outcomes with development processes, thus aiding theory and practice (Wang & Hannafin, 2005).

This study employed a two-level cycle of DBR. The *learner-level* cycle involved experimentation, analysis and redesign done for each learner within the same lesson. In this cycle, the teacher reflected on lessons for each learner and applied improvements as needed. The *lesson-level* cycle involved design, experimentation, and analysis, done after all learners had gone through one lesson or teaching experiment (TE). This cycle usually ends with a retrospective analysis meeting between the teacher and the researchers. With 10 lessons and 9 learners, the study could theoretically undertake up to 90 cycles of design, testing and analysis. Hypothetical learning trajectories (HLTs) guided the design through theoretical assumptions by the researchers resulting from thought experiments supported by previous studies. More specifically, the following are the three initial HLTs used to guide the design of the VR tours:

(1) SI ← SP. Spatial presence (SP) positively influences situational interest (SI).
(2) IF ← SP. Spatial presence (SP) positively influences ideal future-self (IF).
(3) LO ← SP. Spatial presence (SP) positively influences learning outcomes (LO).

Furthermore, the study integrated the presentation, practice and production method, a common language learning strategy, organising lessons into content presentation, practice activities, and production by students creating with the target vocabulary (Pazmiño Vargas et al., 2023).

The study developed a conceptual framework considering the four key variables – immersive capability, spatial presence, situational interest and the ideal future-self – and their effects on motivation and learning outcomes. It is important to note that this study did not directly assess the impact of immersive capability due to learners' use of provided low-cost VR spectacles, which offered the same level of immersive capability for all learners. Figure 1 shows the conceptual framework of the study, including both the model for observing variable interactions and the methodological model. It also shows the connection of these two models by illustrating the influence of improvements implemented in the DBR cycles on learning outcomes and motivation as well as some of the key variables related to VR.

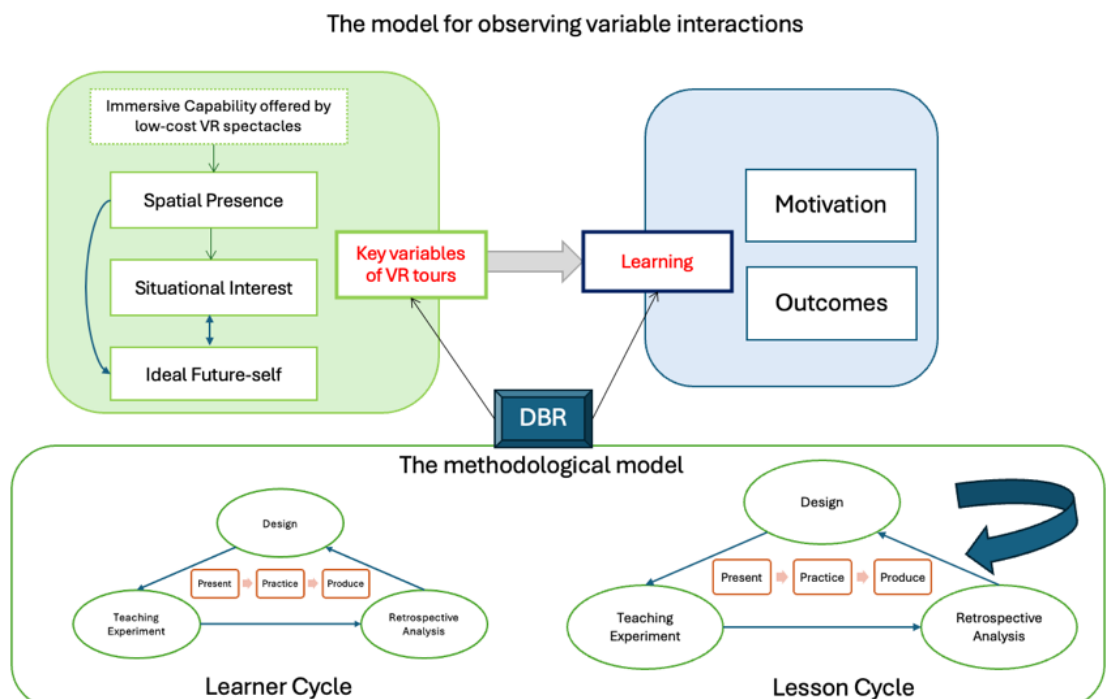

*Figure 1*. Conceptual framework of the study

## Methodology

### Research design and procedure

As mentioned above, the study adopted a DBR design that employed two cyclic levels – learner-level and lesson-level cycles. It was conducted over an extensive period, from April 2019 to December 2019, and followed five phases:
(1) Study and ethical preparation: Prior to the start of the study, the Ethical Review Board of our universities reviewed and approved the research proposal, which adhered to ethical standards such as participant anonymity and informed consent.
(2) Design and development of materials: In collaboration with a language teacher, we designed and developed the initial version of the materials for the 10 lessons.
(3) Lesson-based DBR cycles: The process began with a draft lesson, tested through a TE with participants. After each participant completed the lesson, we analysed the collected data to refine the next lesson and revise theoretical assumptions. We also conducted Interviews after the first, fifth and 10th lessons.
(4) Learner-based DBR cycles: Each lesson-based DBR cycle included a TE for participants with the teacher. We analysed the data from one TE to inform and improve the succeeding TE.
(5) Data analysis: We analysed the collected data through both quantitative and qualitative methods to address the research questions.

### Participants

Participants were selected using purposive sampling based on their interest, perceived need to learn English and willingness to participate. Questions were used to interview and select candidates. Recruitment occurred from May to July 2019, focusing on Japan and France. These two countries, located on different continents, have experienced a growing demand for English due to their role in foreign trade and the globalisation of businesses (Morita, 2017; Salomone, 2022).

Nine students, five French and four Japanese, participated in the study. These learners represented diverse professional, generational and cultural backgrounds and had varied motivations for learning English. In accordance with our confidentiality agreement, pseudonyms are used to describe the participants' characteristics and backgrounds in this section. In a similar vein, "Cher" is the pseudonym for the study's teacher.

Table 1 presents the participants' pseudonyms alongside their background information. Their proficiency in English was assessed using the Common European Framework of Reference for Languages. This estimation was based on the results of an interview and two online tests.

Table 1
*Participants and their background*

| Participant | Level | Location | Occupation | Age | Nationality |
|---|---|---|---|---|---|
| Eiji | B2 | Kyoto, Japan | Graduate student | 24 | Japanese |
| Eito | B1 | Kanagawa, Japan | Retired banker | 70 | Japanese |
| Mizuki | B1 | Chiba, Japan | Cram school manager | 26 | Japanese |
| Masaki | B1 | Tochigi, Japan | Professor | 45 | Japanese |
| Denis | B1 | Grenoble, France | Sales officer | 37 | French |
| Eiman | B1 | Bordeaux, France | Engineer | 32 | French |
| Emma | A2 | Bordeaux, France | Engineer | 34 | French |
| Este | C1 | Paris, France | Architect | 29 | Italian |
| Charles | A2 | Grenoble, France | Engineer | 34 | French |

- Eiji, a mid-20s university exchange student with a background in America, joined the study to try a new language-learning method; he was interested in translating English news to Japanese and following Philippine events.
- Eito, a 70-year-old retired banker from Yokohama, learned English through radio and online classes and joined to improve his vocabulary and conversation skills.
- Mizuki, a cram school manager with a master's degree in education, volunteered to enhance her communication skills, inspired by her experience in Europe.
- Masaki, a psychology professor in Tokyo, volunteered for the new lesson structure because he was interested in the study and technology.
- Denis, living in Grenoble, aims to learn English for better communication with his bilingual daughter and her relatives, joining to expand his vocabulary.
- Emma, an engineering firm employee in Bordeaux, sought a new learning method to fuel her travel and international conversations.
- Charles, a Grenoble-based semiconductor technician, joined the study to improve his English for family vacations.
- Este, an Italian architect in Paris, left after the first class due to his high proficiency.
- Eiman, a Parisian entrepreneur, exited after five classes due to work and travel obligations.

Este and Eiman, both French participants, left early but offered valuable insights for retrospective analysis.

**English learning materials**

TEs were conducted by the teacher for each participant using English learning materials organised into 10 lessons and studied the lessons independently. Each lesson and tour featured five target vocabulary words.

The lesson guide with target vocabulary, exercises, and virtual tours was developed with Cher, a Filipino online English teacher with over 3 years of experience teaching Japanese learners English via Skype and Zoom. Each TE had a presentation, practice and production lesson guide to help students study. The lesson guides followed this structure:
- A pre-lesson vocabulary test featuring the target vocabulary for the lesson
- Presentation: Instructions and a link to the VR tour, including a list of target vocabulary words and their pronunciation
- Practice: Four exercises utilising the target vocabulary, such as reading conversations, fill-in-the-blanks, odd-one-out and matching exercises
- Production: Instructions for using the target vocabulary in constructing their own sentences for the VR tour
- A post-lesson vocabulary test with the lesson's target vocabulary.

The materials for the lessons and other resources are freely available online under the Creative Commons Attribution Non-Commercial Share-Alike 4.0 License (CC-BY-NC-SA) and can be accessed at https://iop.upou.edu.ph/portfolio/projects/.

During class, students used lesson guides that included target vocabulary, exercises and space to write a tour script. Table 2 shows each detail of the learner's time of participation, most-used synchronous communication technology and preferred learning space.

Table 2
*Learners contextual and temporal details*

| Learner | Time | Technology | Learning space |
| --- | --- | --- | --- |
| Eito | Morning or afternoon | Skype (Video) | House, Japan |
| Eiji | Morning | Skype (Video) | Café, Japan |
| Mizuki | Morning | Skype (Video) | House, Japan |
| Masaki | Afternoon | Zoom (Video) | House, Japan |
| Denis | Morning | Facebook Messenger (Video) | House, France |
| Emma | Lunchtime | Facebook Messenger (Audio) | Workplace, France |
| Charles | Lunchtime | Skype (Video) | Workplace, France |

**VR tour and device**

The 10 VR tours, one for each lesson, were progressively developed through the DBR cycles. Each tour featured a 360-photo of a location known for its significant number of English-speaking locals. The vocabulary for each tour, ranging from B1 to C1 levels, was carefully chosen in accordance with the recommendations from the English language profile word list of the Common European Framework of Reference (https://englishprofile.org/wordlists). An example of a tour integrated into Lesson 10 is illustrated in Figure 2. Throughout the study, all 10 VR tours were subject to a series of minor modifications, informed by findings from retrospective analyses.

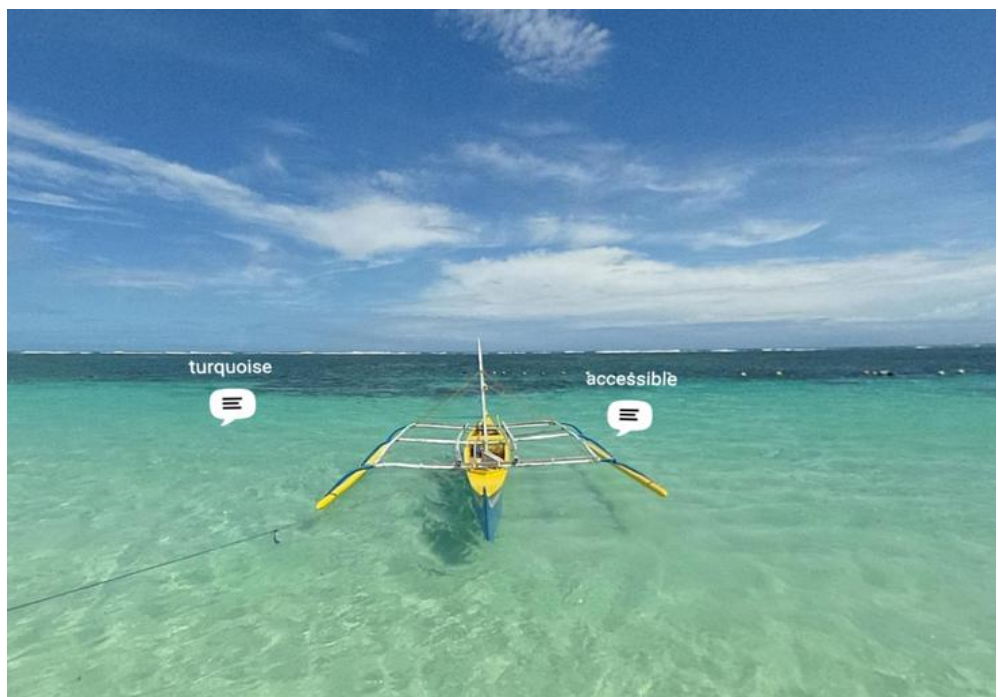

*Figure* 2. An example of VR tours: VR tour used in Lesson 10

All the France-based learners, except for Denis, who had a susceptibility to epileptic shocks, were provided with low-cost VR spectacles for their smartphones to be used as head-mounted displays (HMDs). They received in-person training on how to use the device from the first researcher or an assistant. The learners were instructed to start using the spectacles in the sixth lesson cycle. However, none of them ultimately utilised the spectacles during the TEs for reasons that are elaborated upon in the Findings section.

**Data collection**

To gather a comprehensive data set throughout the study, multiple instruments were utilised, including observation notes, interview protocols, survey questionnaires, teacher reflections and tests.

The teacher and I (RBF) filled out observation notes during and after each TE with the aid of video and audio recordings. A standardised template was used for these notes to ensure focus on key aspects of the experiment and relevant data collection. The notes focused on challenges in the lessons and how they were solved. Learner comments and cues from their behaviour or discussion were also noted. Furthermore, proposed revisions in the lesson were also listed.

Interview data were collected after the first, fifth and last TEs. Structured interview protocols were prepared for these sessions, categorised as initial, midpoint and final interviews. The questions were crafted to draw out the best features of the VR-based lessons as well as the obstacles that students encountered in their learning environment. Questions that helped them describe the image and vividness of their ideal future-self as well as the interesting features of the lessons were also asked.

Survey questionnaires were distributed online after each TE from Lessons 2 to 10, totalling nine surveys. In each survey, learners provided quantitative feedback on spatial presence, situational interest, ideal future-self, overall experience and satisfaction, along with reasons for their ratings. The surveys included 10-point scales for measuring spatial presence, situational interest and ideal future-self, complemented by open-ended questions. The spatial presence item was based on Bouchard et al.'s (2004) validated single-item scale. Situational interest items were derived from Hidi and Renninger's (2006) definitions of triggered and maintained interest. The item for the ideal future-self was adapted from Lamb's (2012) Ideal L2 Future-Self scale.

Additionally, Cher, the teacher, provided her reflections after each session using a feedback guide. These reflections were used in conjunction with our observations for retrospective analysis.

Pre-tests and post-tests consisting of multiple-choice or matching-type questions assessed learners' knowledge of five target vocabulary words in each lesson.

The data collection process is shown in Figure 3. Each TE was recorded, with interviews conducted during the first, fifth and 10th lesson cycles. Surveys following subsequent TEs. Post-lesson data, including spatial presence, ideal future-self and situational interest, were gathered, and score gain was calculated by subtracting pre-test from post-test scores.

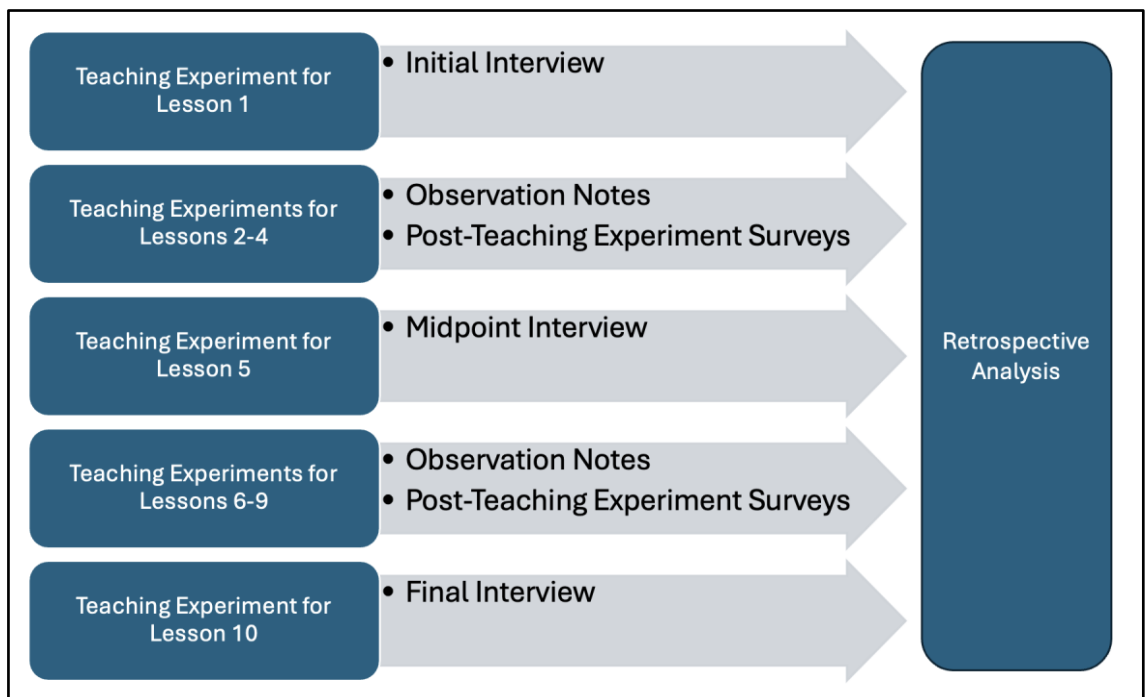

*Figure 3*. Data collection diagram

**Data analysis**

TEs or learner cycles were the subject of reflection on every four or five meetings. During the meetings, Cher, the teacher, reflected on her notes for each student, while we read and discussed the observation notes, interviews and surveys, identifying trends and areas for improvement. This process of retrospective analysis helped refine HLTs and establish action points for future lessons, as depicted in Figure 4.

After all the qualitative data was gathered, a reflexive thematic analysis was conducted, following Braun and Clarke's (2019) method, on answers to the open-ended questions of the nine surveys, the observation notes and the three interviews. First, data familiarisation and organisation were carried out using NVivo. Second, initial codes were generated and used to label chunks of text that were relevant to the research questions. Third, codes were grouped into themes and subthemes. Fourth, the themes and subthemes reviewed and reorganised. Finally, themes and subthemes were defined and organised according to the research questions.

Furthermore, visualisations were generated from variable ratings using R software. Since spatial presence's single-item scale was the only validated measure in the survey, it, along with score gain, was the focus of hypothesis testing. We examined variations in spatial presence and score gain across lessons (1 through 10), age groups (20s, 30s, 40+), cultures (Japanese, French, cosmopolitan) and physical learning environments (home, workplace, café).

For hypothesis testing, a Friedman rank-sum test and a Nemenyi post-hoc test were conducted using the statistics libraries R (R Team, 2021) and Pairwise Multiple Comparison of Mean Ranks (Pohlert, 2014).

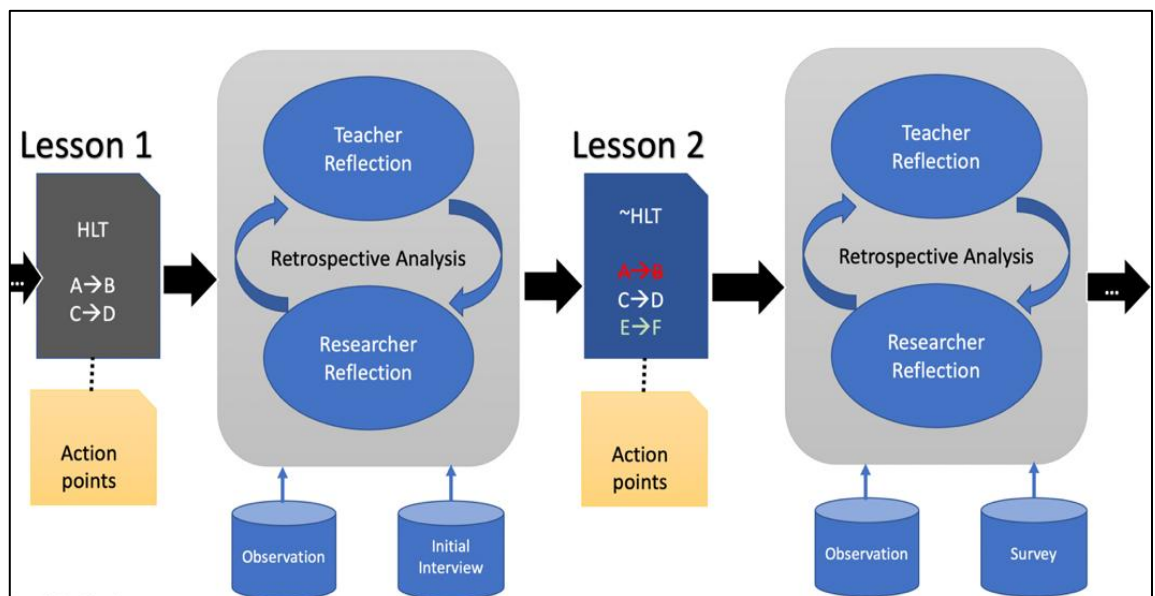

*Figure 4.* Retrospective analysis process from Lesson 1 to Lesson 2

# Findings

This section presents the findings from the retrospective, thematic and quantitative analyses, aligned with the study's objectives.

**RQ1: Learner experiences and lesson improvement**

Retrospective analyses allowed significant improvements to be made in lesson design. In addition to clerical changes such as enhancing the clarity of instructions, correcting typographical errors and making aesthetic enhancements, pedagogical modifications led to the development of new HLTs based on underlying theories. These HLTs were then confirmed in the observation notes and discussions with the teacher in succeeding retrospective analysis meetings. For example, the teacher observed that after adding text-based scaffolds and VR tour transcripts, the learners picked up the vocabulary words more easily from the VR tours. Moreover, misunderstandings regarding some vocabulary words after presenting words from the same part of speech (from Lesson 5) occurred less frequently compared to the earlier lessons.

Table 3 lists these pedagogical additions, their corresponding HLTs and implementation details. These include text-based scaffolds, such as VR tour transcripts, and image-based scaffolds to clarify learning content and enhance word understanding. Another addition was standardising vocabulary presentation and adding retrieval practices from Lesson 6 onwards as they improved students' recall and supported the learning process by addressing part-of-speech mismatches.

Table 3
*Pedagogical revisions*

| Pedagogical revisions | Theoretical revisions | Lesson instances |
| --- | --- | --- |
| Providing text-based scaffolds | LO ← Sc | 2, 3, 4, 5, 8 |
| Providing image-based scaffolds | LO ← Sc | 3, 6 |
| Presenting vocabulary words in the tour that were of the same part of speech (e.g., nouns, adjectives, verbs) | LO ← PS | 4 |
| Reusing words from previous lessons in activities | LO ← RO | 4, 6 |

*Note*. LO: learning outcomes; Sc: scaffolding; PS: part of speech; RO: retrieval opportunities.

### RQ2: Variable dynamics

In this study, dynamics, conceptualised as forces driving change in a system, were operationalised through HLTs. This section summarises these factors, which are organised into themes.

*Factors influencing situational interest*
First, situational interest generally increased throughout the lessons, with a few exceptions (i.e., Masaki and Mizuki). This trend was evident in their responses during both the midpoint and final interviews.

As expected, learners reported that spatial presence significantly influenced their interest. This was supported by 36 instances, where statements such as "I went to the place that was interesting" and "it was enjoyable because [it] was a realistic experience".

Additionally, an aspect of individual interest, specifically the learner's interest in the place featured in the tour, was also found to affect their engagement with the tour. This was supported by 30 instances, where statements such as "I have a dream to go to Canada but not in the marketplace" and "I would like to visit the place, and I would like to stay there for a long time". Masaki's varying interest in the activities depending on the featured location in each VR tour lesson served as a particularly good example of this.

In the thematic analysis, perceived learning increased learners' interest in the activities. This was supported by 26 instances of statements such as "I could enjoy the time to know new vocabulary and things I didn't know" and "My feeling is good because I have learned a lot of words and synonyms".

The novelty of the VR technology played a significant role in situational interest, as many learners enjoyed and were curious about its newness. This was supported by 24 instances of statements such as "It's really exciting that I had such a kind of a VR experience in learning English. Was thinking like VR equipment is only for gaming and any other traveling. However, it's also useful and effective for English learning tool. It was awesome" and "VR tour is very interesting because it's a new method".

Additionally, the unique aspect of VR tours, offering a sense of discovery through new places or scenery, further enhanced their interest. This was supported by 12 instances of statements such as "because I don't know the Philippine hero and I want to know the special area in the foreign country such as the Rizal Monument" and "especially I enjoyed VR tour because I can see sights which I've never been to".

This enjoyment was often linked to the beautiful scenery featured on the tours; supported by 23 instances of statements such as "The positive feelings I had were [the] speaker's excellent pronunciation and beautiful scenery" and "This time the scenery was very beautiful, and the soundtrack was pleasant".

Finally, several students noted that the teacher's qualities, such as her helpful attitude and enthusiasm, contributed to their heightened interest. This was supported by 11 instances of statements such as "The explanation was easy to understand" and "great teacher".

*Factors influencing the ideal future-self*
The dynamics of ideal future-self were less observable in the TEs but emerged in survey and interview analysis. Learners' perceived learning progress influenced their view of themselves as improved English speakers in the future. This was supported by 10 instances of statements such as "I see myself getting better in English in the future because it helps me in getting more vocabulary. I can say I improved a bit" and "I'm sure that my vocabulary is increasing, even though my answer in the pre-test is correct, sometimes I don't know the exact meaning and I am not sure if it's correct or not. Vocabulary helps me".

Another contributing factor to the positive development of learners' ideal future-self was their instrumental motivation for learning English. Some envisioned themselves as tourists conversing with locals in English, while others aimed to use the language professionally or to enhance communication with relatives and international friends. This was evidenced by 49 instances of statements such as "I want to use English in traveling and maybe a little bit at work" and "Discussion with my Australian family".

*Factors influencing spatial presence*
Spatial presence ratings were plotted for each learner in Lessons 2 to 10, as illustrated in Figure 5. A rating of 10 indicates a very high sense of presence while a rating of 1 indicates a very low sense of presence. The plot shows a general increasing trend among most learners, with the exceptions of Mizuki and Masaki (discussed further in the findings of the RQ3: Learner differences section). This upward trend is likely due to improved self-efficacy in later lessons, as they became more adept at navigating the tours and less hindered by earlier technical difficulties. Nevertheless, the Friedman test indicated no statistically significant variation across the lessons ($F(8) = 10.43$, $p = 0.24$).

The impact of immersive capability on spatial presence was limited, as learners did not consistently use HMDs. Some, like Emma and Eiman, found HMDs to be uncomfortable and caused dizziness. Denis could not use it due to his physiological situation, and Eito found the setup distracting, though he noted stronger immersion when reviewing lessons with the HMD. Furthermore, blurred lenses sometimes hindered Eito's experience, and using the HMD within the short lesson duration (30–40 minutes) was inconvenient for him.

Quantitatively, the Friedman sum rank test showed no significant variance in score gain between lessons ($F(8) = 13.69$, $p = 0.09$).

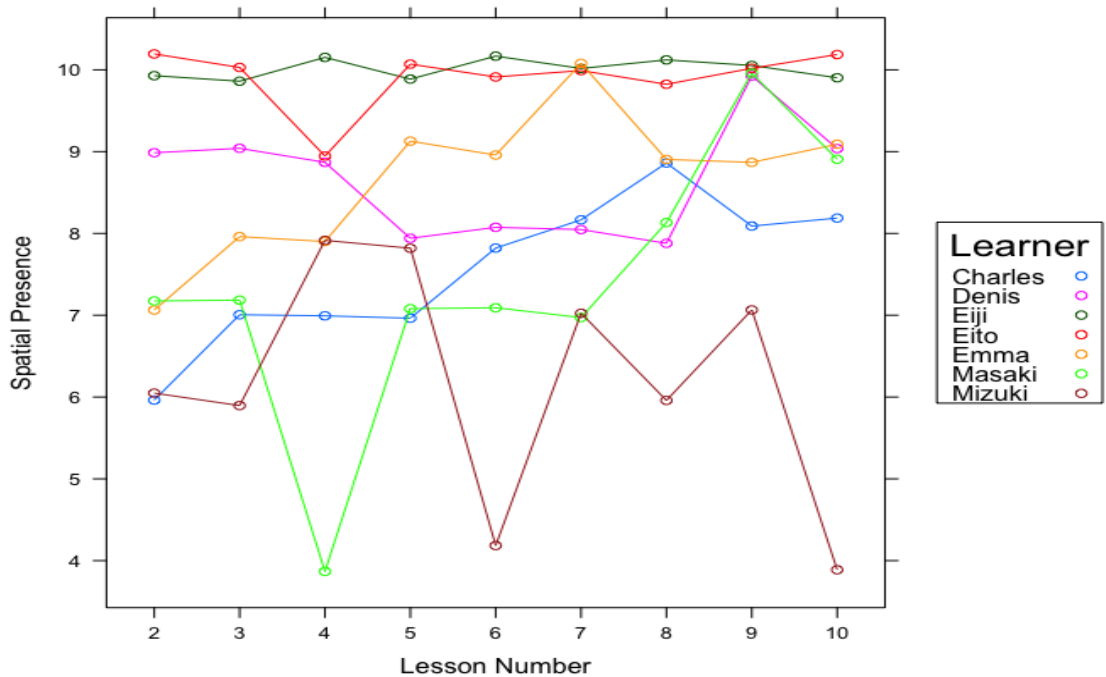

*Figure 5.* Spatial presence experienced by learners across lessons

**RQ3: Learner differences**

Denis and Eiji, having lived in various English-speaking countries, were categorised as cosmopolitan. Other learners were grouped based on their lifelong residences: Masaki, Eito and Mizuki from Japan; and Emma and Charles from France.

Quantitative analysis showed cultural background, age and physical learning environment differences in score gain. The Friedman sign-rank test showed no significant differences in spatial presence by age group ($F(2) = 4.34$, $p = 0.11$) or learning environment ($F(2) = 1.33$, $p = 0.51$), but by culture ($F(2) = 6.61$, $p = 0.04$). Cosmopolitan learners had the highest median spatial presence. Nemenyi post-hoc tests showed no significant group differences. Significant differences in score gains were observed based on learning environment ($F(2) = 10.89$, $p < 0.01$), with differences between home, work and café settings. The Nemenyi post-hoc test revealed significant differences in score gains between home-based and workplace-based learners ($p < 0.05$), as well as between home-based and café-based learners ($p < 0.05$).

Differences in learning styles were also observed: Japanese learners focused more on verbal self-monitoring, French learners made more mistakes but were creative and cosmopolitan learners focused on understanding word usage in different contexts.

Cultural differences did not distinctly affect interest in the VR tours, but individual interest in the featured places did. For instance, Masaki showed more interest in VR tours featuring the Philippines than the United States of America. Eito, an outlier among Japanese learners, exhibited high individual interest in English, maintaining enthusiasm regardless of VR tour quality.

Learners' visions of their ideal future-self varied: cosmopolitan learners envisioned more fluent conversations with family and friends, French learners anticipated travel, while Japanese learners found it harder to picture their ideal future selves.

# Discussion

Research findings align with literature, highlighting that situational interest is boosted by spatial presence, as supported by Cheng (2022) and Figueroa (2023), demonstrating that VR-based learning sustains student motivation (Alizadeh & Cowie, 2022). The increase in spatial presence owing to experience and less technical learning curve obstacles contributed to the favourable trend of situational interest in subsequent classes.

The enhanced visualisation of participants' ideal future selves, influenced by the instrumental benefits of language learning, aligns with findings from Dörnyei (2019). This was primarily concerning the places students aspired to visit, suggesting that virtual exploration of these destinations heightened their future visitation desires, thereby boosting their motivation to learn English. Learning outcomes from this study validated Figueroa's (2023) recent experimental findings that spatial presence enhances learning outcomes. Furthermore, the unexpected findings offered fresh insights.

## Sustained novelty in VR learning

Initially, we anticipated that the novelty of VR as an emerging educational tool and the unique design of VR tour lessons would spark learners' interest, is in line with findings in the literature (Deci, 1992; H. Sun et al., 2008). However, contrary to the common belief that novelty fades with familiarity (Jeno et al., 2019), findings from our study indicated that interest in most of the lessons was positively influenced by the new locations and their beauty. Scenic novelty in the dynamics of spatial presence and interest aligns with findings in tourism studies, where visitors seek unique, distinct destinations from their previous travels (Blomstervik & Olsen, 2022; Niininen et al., 2004). Furthermore, our findings support studies that suggests beauty naturally draws and maintains human attention (Sui & Liu, 2009; Vinh, 2013). This insight is crucial for instructional designers, as it emphasises the importance of carefully selecting visually appealing elements in VR programmes. These elements not only enhance spatial presence but also play a vital role in engaging learners in VR activities.

## VR headset usage explained by adult and non-formal learning theories

A surprising trend was that students who received HMDs for VR tours rarely used them. HMDs increased spatial presence (Witmer & Singer, 1998), but their use in real life differed from controlled environments. One reason students stopped using HMDs was dizziness. This shows that physiological discomfort (Lessiter et al., 2001) can negate spatial presence's benefits. Furthermore, the provided HMDs were low-cost VR spectacles, potentially less comfortable and not fully compatible with the learners' smartphones. Although research comparing high-end HMDs and budget smartphone VR headsets found little difference in spatial presence, usability and cybersickness (Lombard & Ditton, 1997), Papachristos et al. (2017) did note a higher physical demand for users of the low-cost models compared to those using premium devices.

Our study also suggests that non-immersive VR tours might be more suitable for adult learners unfamiliar with VR technology and giving them the option to choose enabled them to maximise the technology for learning as explained by non-formal learning theories like heutagogy (Blaschke, 2012).

However, the situation could differ for VR-experienced adult learners or in-person settings with high-quality HMDs and more adequate support. With advancements in low-cost VR HMDs and growing VR familiarity, future studies may yield more favourable outcomes for these devices.

## Pedagogical improvements in DBR

New HLTs from this DBR study confirmed studies on teaching and learning vocabulary. Pagán and Nation (2019) observed that repeating target vocabulary across exercises enhanced learning, a finding echoed by our findings. Additionally, Barcroft (2015) indicated that increasing opportunities for retrieval improves word knowledge.

The value of DBR was highlighted in our study with the confirmation of initial HLTs while adding new ones in the design and implementation of the lessons through learner feedback and reflective discussions which are very specific to a context (Bakker & Van Eerde, 2015). In the case of our study, the findings from retrospective analyses revealed how adult learners differed in using the VR-based learning material depending on their physical learning environment and schedule. The theories developed and revised from our study may be humble in a sense that they cannot be readily generalised for all other contexts because they were tested in a particular learning environment. However, they can be tested in similar contexts especially with lessons that use tours and exploration as tools for teaching vocabulary in other technology-enhanced environments like the metaverse or virtual worlds.

## Conclusion

In our study, we explored how spatial presence, ideal future-self, situational interest and learning outcomes interacted in a DBR setting. This involved 10 30-minute one-on-one VR tour lessons aimed at teaching English vocabulary to four adult learners in Japan and five in France. Our quantitative, retrospective and thematic analyses largely confirmed the anticipated influence of spatial presence on situational interest, ideal future-self, and perceived learning.

Additionally, relationships with motivational variables that were not anticipated based on previous studies were uncovered, including novelty of the scenery as a significant environmental factor, alongside various learner-specific influence such as the physical environment where learners joined their online lessons. Contrasting expected outcomes, immersive capability did not play an important role in increasing spatial presence as the HMDs were not convenient or comfortable for the learners in this context. Finally, relationships between lesson improvements applied in succeeding lessons and perceived learning and situational interest established the value of DBR not just as a research approach but also as a reflective way of developing effective learning interventions.

Although our study provides insightful observations, it also presents limitations that highlight areas for further research. Our findings may be difficult to generalise because of the number of participants and our methodological approach. However, the study as well as the pedagogical strategies gained from it can be replicated in similar contexts for further validation and refinement.

In conclusion, our study offers valuable insights for both academic understanding and practical application of using VR tours in online language learning. Furthermore, it sheds light on the interaction and influence of technological and scenic novelty on situational interest, especially for learners repeatedly exposed to VR tour lessons. From a practical standpoint, the lessons drawn from our study are poised to inform and assist online educators and instructional designers in effectively incorporating VR-based activities into their online lessons. This could enhance the overall learning experience and engagement of students in a virtual learning environment.

## Author contributions

## Acknowledgments

We extend our deepest gratitude to the Japan International Christian University Foundation for their generous support and funding of this research project.

This paper was refined with the assistance of OpenAI's GPT-4o and QuillBot, complementing our editorial process.

---

**Corresponding author:** Roberto B. Figueroa Jr., robertojr.figueroa@up.edu.ph